\documentclass[12pt]{iopart}
\usepackage{iopams}
\usepackage{graphicx}

\begin{document}
\newcommand{\bra}[1]{\left\langle #1 \right|}
\newcommand{\ket}[1]{\left| #1 \right\rangle}
\title[Diffusion on the interval]{Diffusion on the interval acted upon by a oscillating space-homogeneous force}
\author{Ev\v{z}en \v{S}ubrt\footnote{corresponding author} and Petr Chvosta}
\address{Department of Macromolecular Physics, Faculty of Mathematics and Physics, Charles~University,
V~Hole\v{s}ovi\v{c}k\'ach~2, CZ--180~00~Prague, Czech Republic}
\eads{\mailto{subrt@kmf.troja.mff.cuni.cz}}
%
%
\begin{abstract}
We study the one-dimensional diffusion process which takes place between two reflecting boundaries and which is acted upon by a time-dependent and spatially-constant force. The assumed force possesses both the harmonically oscillating and the constant component. Using techniques presented in the paper~\cite{chvosta05b} we derive the set of integral equations whose comprise the Green function for our diffusion process. In the later part we focus on the time-asymptotic stationary regime of considered non-equilibrium process. Some of its time-asymptotic characteristics, e.g.\ time-averaged and time-asymptotic probability density of the particle coordinate, exhibit nontrivial features.
\end{abstract}
%

%
\section{Introduction}
Consider a rectangular vessel filled with small Brownian particles dissolved in a good solvent. With no external force acting on our particles, the concentration of the particles would be homogeneous. Now assume that each particle carries the same electric charge. What would be the concentration profile if we put the vessel in an external, constant and homogeneous electric filed? After some time it approaches the new constant equilibrium profile. Its shape would depend on many factors, like temperature of the solvent, dimensions of the vessel, mobility of the particles and a strength of the field. And what if the intensity of the filed would oscillate?

We assume that the external homogeneous field acts only along the one axis of the vessel, let say along the $x$-coordinate, that the particles do not interact between themselves and that the concentration of the particles is homogeneous in the other ($y$ and $z$) directions. Under these circumstances we can treat the emerging dynamics as a one-dimensional diffusion process confined to the spatial interval by two reflecting boundaries and acted upon by a space-homogeneous and time-dependent force. 

Since the beginning of the $20^{\mathrm{th}}$century the diffusion problems concerning the movement of the Brownian particle in a fixed potential profile have been investigated many times in the literature and nowadays it is a well understood problem \cite{vankampen92, gardiner85, risken96}. In a past few decades the diffusion dynamics was reexamined in the systems, in which the potential incorporates a time-dependent component. Notably interesting phenomena such as a stochastic resonance \cite{gammaitoni98, anishchenko02} or effect of Brownian motor \cite{reimann01} arise if the potential changes periodically in time. It is known experimentally that nonlinear effects of alternating electric field are very important in governing the function of membrane proteins \cite{astumian89, liu90}.

The time-dependency of the potential represents the fundamental ma\-the\-ma\-ti\-cal difficulty in the theoretical treatment of the diffusion problem. Nowadays there are only few exactly solvable diffusion problems with the time-dependent potential \cite{hugang90, chvosta03a, chvosta03b, subrt06}. Moreover, the presence of a (reflecting/absorbing) boundary rises further the complexity of the computations.

The next section presents the general treatment of the problem based on the computational techniques presented in~\cite{chvosta05b}. In section~3 we assume that the force, aside from its constant part, incorporates the harmonically oscillating component. Section~4 presents the procedure, which allows the improved discussion of the time-asymptotic dynamics. Section~5 is focused on the time and space resolved probability density and its period-averaged counterpart. Section~6 contains the concluding remarks and summarizes the results.
\section{Green function for a diffusion between two boundaries}
Consider a free diffusion of a Brownian particle under the influence of a space-homogeneous and time-dependent force
$F(t)$. Such setting corresponds to the time-dependent potential of the form $V(x,t)=-xF(t)$. In the overdamped regime, the
time-evolution of the probability density for the position of the diffusing particle is determined
by the Smoluchowski equation \cite{risken96}
\begin{equation}\label{smoluchowski}
  \frac{\partial}{\partial t}p(x,t)=-\frac{\partial}{\partial x}\left[-D\frac{\partial}{\partial x}p(x,t)+v(t)p(x,t)\right]\,\,.
\end{equation}
Here, $v(t)$ is the time-dependent velocity of the drift motion, which is induced by the external
force, $v(t) = F(t)/\Gamma$. Parameter $\Gamma$ equals the particle mass times the viscous friction coefficient. Its reciprocal
value plays the role of the particle mobility. The thermal-noise strength parameter $D$ is the linear function of the absolute temperature, $D = k_{\rm{B}}T/\Gamma$. The expression in the square brackets on the right-hand side of \eref{smoluchowski} represents the probability current.

The solution of Smoluchowski equation \eref{smoluchowski} with the assumed time-dependent potential is accessible using the various algebraic methods \cite{wolf88, suzuki85} reads
\begin{equation}\label{g0}
  G(x,t;x',t')=\frac{1}{\sqrt{4D\pi(t-t')}}\exp\left\{-\frac{\left[x-x'-\int_{t'}^t\rmd t''v(t'')\right]^2}{4D(t-t')}\right\}\,\,.
\end{equation}
This Green function yields the solution of \eref{smoluchowski} for the initial condition at time $t=t'$ in the form $\lim_{t\to t'}p(x,t)=\delta(x-x')$. 
Qualitatively, equation \eref{g0} represents the gradually spreading Gauss curve, whose center is located at the coordinate $x=x'+\int_{t'}^t\rmd t''v(t'')$.

Now assume, that there are two reflecting boundaries located at $x_1$ and $x_2$, with $x_1<x_2$. Let the particle is initially placed at the arbitrary coordinate between the boundaries, i.e.\ $x'\in(x_1,x_2)$. Without loss of generality we set the time origin at $t=0$. The resulting diffusion dynamics is restricted by the boundaries to the space-interval of length $l=x_2-x_1$. For the Green function of the present problem we assume the form
\begin{eqnarray}\label{ansatz}
	U(x,t;x',0)=G(x,t;x',0)-D&\int_0^t\rmd t'\frac{\partial}{\partial x}G(x,t;x_1,t')B_1(t',x')+\nonumber\\
	  &+D\int_0^t\rmd t'\frac{\partial}{\partial x}G(x,t;x_2,t')B_2(t',x')\,\,,
\end{eqnarray}
for all $x,x'\in(x_1,x_2)$ and $t\geq0$. Mathematically, our diffusion problem reduces to the determination of two unknown functions $B_1(t,x')$ and $B_2(t,x')$.

The next steps follow the procedure that was presented in great mathematical details in \cite{chvosta05b}. The reader interested in the mathematical subtleties is referred to the mentioned paper. The requirement of vanishing probability current through every reflecting boundary, together with our ansatz \eref{ansatz}, leads to the set of two Volterra integral equations of the first kind for the unknown functions $B_1(t,x')$ and $B_2(t,x')$,

\begin{equation}\label{ieq1}
\fl D\!\!\int_0^t\!\!\rmd t'\!\!
\left[\begin{array}{r r}
	G(x_1,t;x_1,t')&-G(x_2,t;x_1,t')\\-G(x_2,t;x_1,t')&G(x_2,t;x_2,t')
\end{array}\right]\!\!
\left[\begin{array}{c}
	B_1(t',x')\\B_2(t',x')
\end{array}\right]\!\!=\!\!
\left[\begin{array}{c}
	\int_{-\infty}^{x_1}\rmd x\,G(x,t;x',0)\\
	\int_{x_2}^{\infty}\rmd x\,G(x,t;x',0)
\end{array}\right]\,.
\end{equation}

\noindent It can be shown, that both solutions have a specific physical meaning. Each one represents the actual probability density at the respective boundary, i.e.\ $B_1(t,x')=U(x_1,t;x',0)$ and $B_2(t,x')=U(x_2,t;x',0)$.

The solutions of the system \eref{ieq1} yields the complete description of the emerging diffusion dynamics. From them one can from in derive all one-time characteristics of our model. This general result holds for an arbitrary time dependency of the potential.
\section{Harmonically oscillating force}
In this preparatory section we concertize the general results of the preceding paragraphs to the specific time-protocol of the potential force $F(t)$. We assume that it oscillates harmonically with a time-independent amplitude $F_1\geq0$ and frequency $\omega$ around a fixed value $F_0>0$. The cases with $F_0\leq0$ and/or $F_1<0$ are easily accessible from the model symmetries. Hence the force induced drift velocity is prescribed by the formula
\begin{equation}\label{driving}
v(t)=v_0+v_1\sin(\omega t)\,\,,
\end{equation}
where $v_0=F_0/\Gamma$ and $v_1=F_1/\Gamma$ respectively. To make the following calculations more transparent, we introduce the substitution $\tau=|v_0|^2t/(4D)$. The formula \eref{g0} turns to
\begin{eqnarray}\label{G0red}
  \fl\nonumber G(x,t;x',t')=&\frac{v_0}{2D}\,\frac{1}{\sqrt{4\pi(\tau-\tau')}}\times\\
  &\times\exp\left\{-\frac{1}{\tau-\tau'}
  \left[(\tau-\tau')-\frac{\kappa}{2}\left[\cos(\theta\tau)-\cos(\theta\tau')\right]
  -\frac{\lambda}{2}(\xi-\xi')\right]^2\right\},
\end{eqnarray}
with $\theta=4D\omega/v_0^2$, $\kappa=v_0v_1/(2D\omega)$, $\lambda=v_0l/(2D)$, $\xi=x/l$ and $l=x_2-x_1$. Here we assume $\omega>0$ and $D>0$.

For a moment, we focus on the case of time-independent force, i.e.\ we put $v_1\equiv0$ in \eref{driving}. Integrals in \eref{ansatz} and \eref{ieq1} now depend on time only through the difference $t-t'$. Therefore the complete solution can be readily obtained using the Laplace-transform method or using the expansion into the eigenfunctions of the evolution equation \cite{sweet70}. The probability density $U(x,t;x',0)$ then relaxes to the equilibrium distribution
\begin{equation}\label{Ueq}
U^{(eq)}(x)=\frac{v_0}{D}\frac{\exp[-\frac{v_0}{D}(x_2-x)]}{[1-\exp(-2\lambda)]}\,\,,\ \ \mathrm{for}\ \ x\in\langle x_1,x_2\rangle\,\,.
\end{equation}
Through this formula, the asymptotic values of the probability density at the boundaries $B_1^{(eq)}$ and $B_2^{(eq)}$, and hence the asymptotic solutions of the system \eref{ieq1} are readily accessible. Specifically
\begin{equation}\label{asympt0}
 B_1^{(eq)}=\frac{v_0}{D}\,\frac{\rme^{-\lambda}}{\rme^\lambda-\rme^{-\lambda}}\qquad\mathrm{and}\qquad
 B_2^{(eq)}=\frac{v_0}{D}\,\frac{\rme^{\lambda}}{\rme^\lambda-\rme^{-\lambda}}\,\,.
\end{equation}
These formulas can be used to obtain the solution in the term of an \emph{adiabatic approximation} \cite{parrondo98, usmani02}. Simply changing $v_0$ to $v(t)=v_0+\sin(\omega t)$ and $\lambda$ to $\lambda(t)=v(t)l/(2D)$ in \eref{asympt0} the adiabatic solutions $B_1^{(ad)}(t)$ and $B_2^{(ad)}(t)$ are obtained. However, such approximation is suitable only for high temperatures, low frequencies and weak driving force $F_1$, i.e.\ if the system can nearly instantaneously relaxes to the equilibrium state corresponding to the actual potential profile.

Now we return to the original setting $F_1\neq0$ and $\omega\neq0$. In this case, the equilibrium distribution $U^{(eq)}(x)$ is never reached. The system asymptotically approaches the non-equilibrium steady state. The system integral equations \eref{ieq1} now represents the principal analytical difficulty. The integrals have no more the convolution structure and the kernels display the weak singularity at the upper integration limit. Regardless we have developed analytical methods, which allow us to make some statements about the solution. These results will be presented in the later sections.

For the numerical solution of \eref{ieq1} one can use the \emph{product integration} method \cite{weiss72, linz69, scharf91}, which treats properly the weakly singular kernels. Its outcome incorporates both the transitory effect and the stationary regime. Since we are interested \emph{only} in the asymptotic steady-state and we want to have the better analytical insight into the long-time behavior, we use another approach.
\section{Long--time limit of the densities at the boundaries}
In this section we present the analytical method for solving the system of integral equations \eref{ieq1}.
Though we cannot yet write the exact expression for its solutions $B_i(t,x')$, $i=1,2$, our method allows us to gather more information of the solutions, mainly about its time-asymptotic behavior.

The system of integral equations \eref{ieq1} represents the fundamental difficulty in the analytical treatment of the emerging diffusion dynamics. The non-convolution structure of the equations rules out a direct application of the Laplace transformation. Another difficulty represents the second power of the nominator in the exponential which appears in \eref{g0}. We got rid of it only at the cost of introducing the Fourier transformation into our calculation. Specifically, we use the integral expression for the function \eref{G0red}
\begin{eqnarray}\label{Gvu}\fl
  G(x,t;x',t')=\\
  \nonumber\fl=\frac{v_0}{4\pi D}\int_{-\infty}^{\infty}\rmd u
  \exp\left\{-(u^2-2\rmi u)(\tau-\tau')-\rmi u\kappa[\cos(\theta\tau)-\cos(\theta\tau')]-\rmi u\frac{v_0}{2D}(x-x')\right\}\,.
\end{eqnarray}

As a next step we expand the exponentials of the cosine into the sums of modified Bessel functions $\mathrm{I}_k(\alpha)$,
\begin{equation}\label{rozklad1}
  \rme^{-\rmi u\kappa\cos(\theta\tau)}=\sum_{k=-\infty}^{\infty}\mathrm{I}_k(-\rmi u\kappa)\rme^{\rmi k\theta\tau}
  \,\,,\quad
  \rme^{\rmi u\kappa\cos(\theta\tau')}=\sum_{k=-\infty}^{\infty}\mathrm{I}_k(\rmi u\kappa)\rme^{\rmi k\theta\tau'}\,\,.
\end{equation}

Now we turn our attention to the right-hand sides of the integral equations. The right-hand side of the first equation in \eref{ieq1}, let designate it $r_1(t,x')$, represents the total probability of finding the particle left from $x_1$, i.e.\ left from the left boundary, in the diffusion problem with the same potential but without any boundary. Analogously, the right-hand side of the second equation, $r_2(x',t)$, represents, the total probability of finding the particle to the right from the right boundary placed at $x_2$. In out case with no boundary and $v_0>0$ the particle gradually tends to the plus infinity. Hence asymptotically the particle will be surely found the right from the considered interval. In other words, the integral on the right-hand side of the first equation in \eref{ieq1} asymptotically vanishes. At the same time the right-hand side of the second equation approaches the unity. According to this observation, we expect the right-hand sides to have a form
\begin{equation}
  r_1(t,x')=\widetilde{r}_1(t,x')\,\,,\qquad r_2(x',t)=1+\widetilde{r}_2(t,x')\,\,.
\end{equation}
where $\widetilde{r}_i(t)$ are the transitory parts, that have the important property $\lim_{t\to\infty}\widetilde{r}_i(x',t)=0$, $i=1,2$, for any $x'$. 

From the assumed form of the right-hand sides, we induce the ansatz for the solutions $B_i(t,x')$, $i=1,2$, i.e.\ we assume
\begin{eqnarray}
  B_i(t,x')=\frac{v_0}{D}\left[\beta_i^{(a)}(t)+\widetilde\beta_i(t,x')\right]\,\,,\qquad i=1,2\,\,.
\end{eqnarray}
Again, we suppose that the transitory parts $\widetilde{\beta}_i(t,x')$ vanish identically for $t\to\infty$. Moreover, from the assumption, that the asymptotic solutions are the periodic function of time with the fundamental frequency $\omega$, we expect the asymptotic parts to have a form
\begin{equation}\label{frozklad}
  \beta_i^{(a)}(t)=\sum_{k=-\infty}^{\infty}\beta_{i,k}\exp(-\rmi k\omega t)\,\,,\qquad i=1,2\,\,.
\end{equation}
Here $\beta_{i,k}$ are the unknown complex numbers and we refer to them as to the \emph{complex amplitudes}.

After all above described preparative steps, we can treat the original system \eref{ieq1} with the Laplace transformation. Using the fact, that the exponentials from \eref{rozklad1} and \eref{frozklad} induce only the shift in the Laplace variable, the result yields
\begin{eqnarray}
  \fl\phantom{-}\frac{1}{\pi}\int_{-\infty}^{\infty}\!\!\rmd u\sum_{m,n=-\infty}^{\infty}
  \frac{\mathrm{I}_m(-\rmi u\kappa)\mathrm{I}_n(\rmi u\kappa)}
  {[u-u_+(z_m)][u-u_-(z_m)]}[\beta_1^{(a)}(z_{m+n})+\widetilde\beta_1(z_{m+n},x')]-\label{eq1}\\
  \fl-\frac{1}{\pi}\int_{-\infty}^{\infty}\!\!\rmd u\sum_{m,n=-\infty}^{\infty}
  \frac{\mathrm{I}_m(-\rmi u\kappa)\mathrm{I}_n(\rmi u\kappa)\,\rme^{\rmi u\lambda}}
  {[u-u_+(z_m)][u-u_-(z_m)]}[\beta^{(a)}(z_{m+n})+\widetilde\beta(z_{m+n},x')]=
  \widetilde{r}_1(z,x')\,\,,\nonumber\\
  \fl-\frac{1}{\pi}\int_{-\infty}^{\infty}\!\!\rmd u\sum_{m,n=-\infty}^{\infty}
  \frac{\mathrm{I}_m(-\rmi u\kappa)\mathrm{I}_n(\rmi u\kappa)\,\rme^{-\rmi u\lambda}}
  {[u-u_+(z_m)][u-u_-(z_m)]}[\beta^{(a)}(z_{m+n})+\widetilde\beta(z_{m+n},x')]+\label{eq2}\\
  \fl+\frac{1}{\pi}\int_{-\infty}^{\infty}\!\!\rmd u\sum_{m,n=-\infty}^{\infty}
  \frac{\mathrm{I}_m(-\rmi u\kappa)\mathrm{I}_n(\rmi u\kappa)}
  {[u-u_+(z_m)][u-u_-(z_m)]}[\beta^{(a)}(z_{m+n})+\widetilde{\beta}(z_{m+n},x')]
  =\frac{1}{z}+\widetilde{r}_2(z,x')\,\,.\nonumber
\end{eqnarray}
Here we introduced the abbreviations $u_{\pm}(z_m)=\pm\rmi[\sqrt{1+z_m}\pm1]$ and $z_m=z-\rmi m\theta$.

In the above equations we now proceed with the $u$-integration using the residue theorem. For the equation \eref{eq1} the integration path has to be closed in the upper half of the complex plain. The only singularities there lies in $u_+(z_{m+n})$. Contrary to this, the integration paths in \eref{eq2} must be closed in the lower complex half-plane and the relevant singularities are located at $u_-(z_{m+n})$.

Now we carry out the time-asymptotic limit in the $z$-domain. We multiply both sides of the equations by the Laplace variable $z$ and perform the limit $z\to0^+$. According to our assumptions, the functions $\widetilde{r}_i(z,x')$ and $\widetilde{\beta}_i(z,x')$, $i=1,2$, have no pole at the origin and they vanish in the considered limit. The only other pole at $z=0$ is incorporated in the function $\beta_i^{(a)}(z_{m+n})$ and according to \eref{frozklad} we can write $\lim_{z\to0^+}z\beta_i^{(a)}(z_{m+n})=\beta_{i,m+n}$ for $i=1,2$. 

The resulting equations read
\begin{eqnarray}
  \fl\phantom{-}\sum_{m,n=-\infty}^{\infty}\frac{\mathrm{I}_m(\kappa\alpha^+_m)\mathrm{I}_n(-\kappa\alpha^+_m)}
  {\sqrt{1-\rmi m\theta}}\beta_{1,m+n}-
  \sum_{m,n=-\infty}^{\infty}\rme^{-\lambda\alpha^+_m}
  \frac{\mathrm{I}_m(\kappa\alpha^+_m)\mathrm{I}_n(-\kappa\alpha^+_m)}
  {\sqrt{1-\rmi m\theta}}\beta_{2,m+n}=0\,\,,\\
   \fl-\sum_{m,n=-\infty}^{\infty}\rme^{-\lambda\alpha^-_m}\frac{\mathrm{I}_m(-\kappa\alpha^-_m)
   \mathrm{I}_n(\kappa\alpha^-_m)}{\sqrt{1-\rmi m\theta}}\beta_{1,m+n}+
  \sum_{m,n=-\infty}^{\infty}\frac{\mathrm{I}_m(-\kappa\alpha^-_m)\mathrm{I}_n(\kappa\alpha^-_m)}
  {\sqrt{1-\rmi m\theta}}\beta_{2,m+n}=1\,\,.
\end{eqnarray}
Here $\alpha_m^\pm=\sqrt{1-\rmi m\theta}\pm1$. The above rather cumbersome expressions can be transformed into the compact matrix form
\begin{eqnarray}
  \phantom{-}\mathbb{L}_1\mathbb{M}_0\mathbb{R}_1&\ket{\beta_1}-
  \mathbb{L}_1\mathbb{M}_{1}\mathbb{R}_1&\ket{\beta_2}=0\ket{0}\,\,,\\
  -\mathbb{L}_2\mathbb{M}_{2}\mathbb{R}_2&\ket{\beta_1}+
  \mathbb{L}_2\mathbb{M}_0\mathbb{R}_2&\ket{\beta_2}=1\ket{0}\,\,.
\end{eqnarray}
The matrix elements of $\mathbb{L}_i$ and $\mathbb{R}_i$ are build up from the Bessel functions,
\begin{eqnarray}\label{Lmat}
	&\bra{m}\mathbb{L}_{1}\ket{n}=\mathrm{I}_{|m-n|}(\kappa\alpha^+_n)\,\,,\qquad
	&\bra{m}\mathbb{L}_{2}\ket{n}=\mathrm{I}_{|m-n|}(-\kappa\alpha^-_n)\,\,,\\\label{Rmat}
	&\bra{m}\mathbb{R}_{1}\ket{n}=\mathrm{I}_{|m-n|}(-\kappa\alpha^+_m)\,\,,\qquad
	&\bra{m}\mathbb{R}_{2}\ket{n}=\mathrm{I}_{|m-n|}(\kappa\alpha^-_m)\,\,.
\end{eqnarray}
and the diagonal matrices $\mathbb{M}_i$ are defined as
\begin{eqnarray}\label{mat1}
	\fl\bra{m}\mathbb{M}_{0}\ket{n}=\frac{\delta_{mn}}{\sqrt{1-\rmi m\theta}}\,\,,\quad
	\bra{m}\mathbb{M}_{1}\ket{n}=\frac{\delta_{mn}\,\rme^{-\lambda\alpha^+_m}}{\sqrt{1-\rmi m\theta}}
	\,\,,\quad
	\bra{m}\mathbb{M}_{2}\ket{n}=\frac{\delta_{mn}\,\rme^{-\lambda\alpha^-_m}}{\sqrt{1-\rmi m\theta}}
	\,\,.
\end{eqnarray}
The detailed analysis of elements of matrices $\mathbb{L}_i$ and $\mathbb{M}_j$ reveals the possibility to rearrange \eref{mat1} into the simpler form
\begin{eqnarray}
  &\mathbb{R}_1\ket{\beta_1}-\mathbb{D}_1&\mathbb{R}_1\ket{\beta_2}=0\ket{0}\,\,,\label{final1}\\
  -\mathbb{D}_2&\mathbb{R}_2\ket{\beta_1}+&\mathbb{R}_2\ket{\beta_2}=1\ket{0}\,\,.\label{final2}
\end{eqnarray}
with 
\begin{equation}
\bra{m}\mathbb{D}_1\ket{n}=\delta_{mn}\,\rme^{-\lambda\alpha^+_m}\,\,,\qquad \bra{m}\mathbb{D}_2\ket{n}=\delta_{mn}\,\rme^{-\lambda\alpha^-_m}\,\,.
\end{equation}
At present time we cannot write the explicit solutions $\ket{\beta_1}$, $\ket{\beta_2}$ of the system \eref{final1}--\eref{final2}. Is shows up, that the complex amplitudes $\beta_{i,k}$ and $\beta_{i,-k}$ are complex conjugated and $\beta_{i,0}$ is always real and positive. The solution can be written in the form
\begin{equation}\label{solution}\fl
  \beta_i^{(a)}(t)=\beta_{i,0}+2\sum_{k=1}^{\infty}\beta_{i,k}\cos(k\omega t-\phi_{i,k})\,\,,\quad
  \phi_{i,k}=\arctan\left(\frac{\mathrm{Re}[\beta_{i,k}]}{\mathrm{Im}[\beta_{i,k}]}\right)\,\,,\quad
  i=1,2\,\,.
\end{equation}
Hence the asymptotic solution of the original system of integral equations \eref{ieq1} are $B_i^{(a)}(t)=\frac{v_0}{D}\beta_i^{(a)}(t)$, $i=1,2$.

The formula \eref{solution} directly imply, that the period-averaged time-asymptotic density at the left or the right boundary is $B_1^{(av)}\equiv\frac{v_0}{D}\beta_{1,0}$ or $B_2^{(av)}\equiv\frac{v_0}{D}\beta_{2,0}$ respectively. It can be proved that there is a tight connection between these values. After multiplying the both equations of the system \eref{final1}--\eref{final2} by the vector $\bra{0}$ from the left and after evoking the properties of matrices involved, one equation reduces to the simple statement $\beta_{2,0}-\beta_{1,0}=1$. Differently speaking, the difference between time-asymptotic and time-averaged values of the density at the right and the left boundary always equals to the ratio $v_0/D$.

The fundamental difficulty in the calculation of the \emph{complex amplitude}s is the complicated inversion of the matrices $\mathbb{R}_1$ and $\mathbb{R}_2$. We have solved the system \eref{final1}--\eref{final2} numerically and inserted the outcome into \eref{solution}. The results are in the complete agreement with the later-time outcome of the product integration method. 

\section{Time-asymptotic probability density}
\begin{figure}
	\centering
		\includegraphics[width=1.00\textwidth]{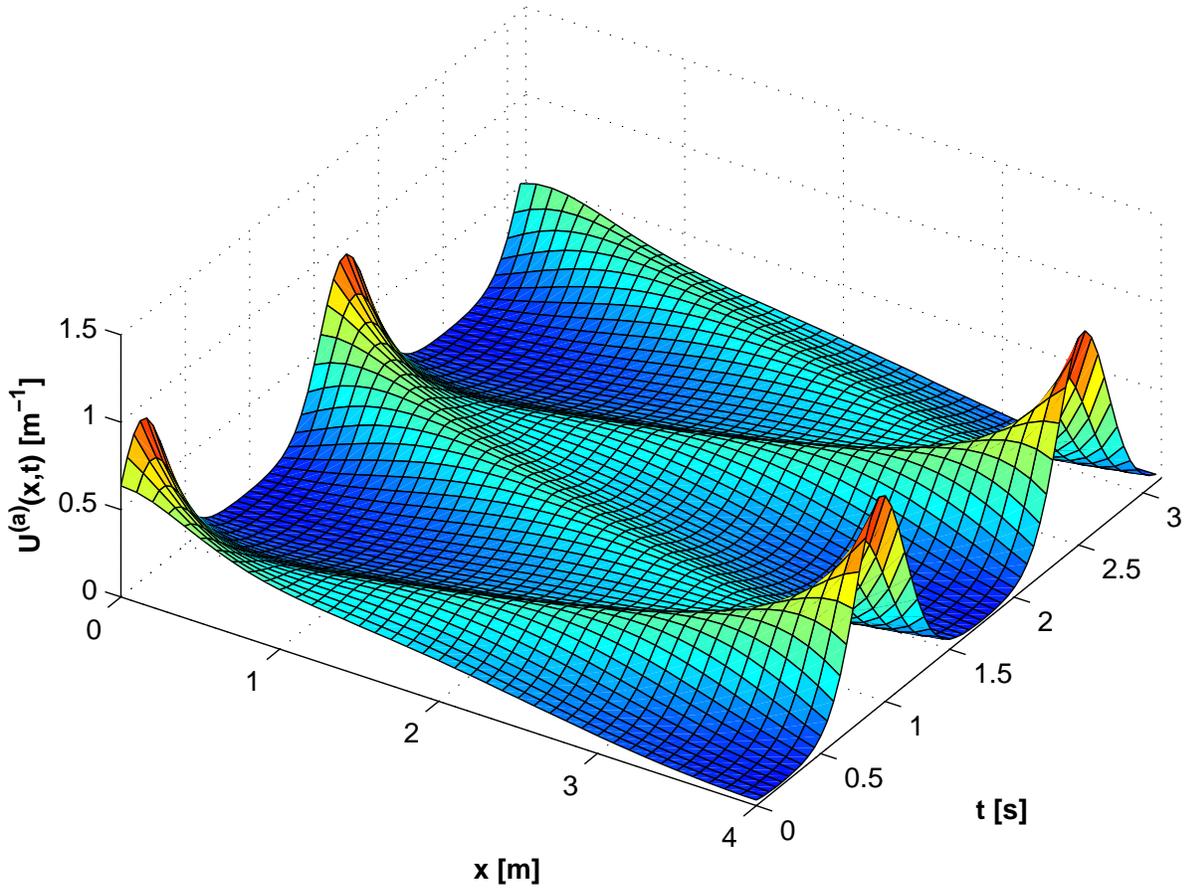}
	\caption{The time and space resolved probability density in the asymptotic regime plotted over two periods of external field. Parameters used are $v_0=0.1\,\mathrm{m\,s^{-1}}$, $v_1=4.0\,\mathrm{m\,s^{-1}}$, $l=4.0\,\mathrm{m}$, $\omega=4.0\,\mathrm{rad\,s^{-1}}$ and $D=1.0\,\mathrm{m^2s^{-1}}$.}
	\label{fig1}
\end{figure}
Assume the vectors of complex amplitudes $\ket{\beta_1}$ and $\ket{\beta_2}$ are known. Now we turn out attention to the asymptotic behavior of the whole probability density, i.e.\ we focus on the function $U(x,t;x',0)\approx U^{(a)}(x,t)$ for $t\to\infty$. In a tight analogy with the ansatz for the densities at the boundaries \eref{frozklad} we expect $U^{(a)}(x,t)$ in the form of its Fourier series
\begin{equation}\label{urozklad}
U^{(a)}(x,t)=\frac{v_0}{D}\sum_{k=-\infty}^{+\infty}\upsilon_k(x)\exp(-\rmi k\omega t)\,\,.
\end{equation}
Of course, it must hold that $U^{(a)}(x_1,t)=B_1^{(a)}(t)$ and $U^{(a)}(x_2,t)=B_2^{(a)}(t)$. To determinate the $x$-dependent Fourier coefficients in \eref{urozklad} we use the similar ``tricks'' as for the calculation of complex amplitudes. We start with the equation \eref{ansatz}. The first term on the right hand side, $G(x,t;x',0)$, for a fixed $x$ represents the gradually diminishing oscillations and vanishes as $t\to\infty$. In the other two terms, we use the expression \eref{Gvu} for $G(x,t;x_i,t')$ and proceed with the $x$-derivation. From this point the procedure is the same as in the preceding section, only the proper integration contours in the complex plain must be chosen during the $u$-integration. Skipping the computational details, the Fourier coefficients $\upsilon_k(x)=\langle k\ket{\upsilon}$ are given by the formula
\begin{equation}\label{upsilon}
	\ket{\upsilon}=\mathbb{L}_2\mathbb{N}_1\mathbb{R}_2\ket{\beta_1}+\mathbb{L}_1\mathbb{N}_2\mathbb{R}_1\ket{\beta_2}\,\,.
\end{equation}
The $x$-dependence of coefficients stems from the matrices $\mathbb{N}_i$. Their elements are given by the expressions
\begin{eqnarray}\label{Nmat1}
  \bra{m}\mathbb{N}_{1}\ket{n}=\frac{\delta_{mn}}{2}\left[\frac{1}{\sqrt{1-\rmi m\theta}}-1\right]
	\exp\left[-\frac{v_0}{2D}(x-x_1)\alpha_m^-\right]\,\,,\\
	\bra{m}\mathbb{N}_{2}\ket{n}=\frac{\delta_{mn}}{2}\left[\frac{1}{\sqrt{1-\rmi m\theta}}+1\right]
	\exp\left[-\frac{v_0}{2D}(x_2-x)\alpha_m^+\right]\,\,.
\end{eqnarray}
The matrices $\mathbb{L}_i$, $\mathbb{R}_i$ and the combinations $\alpha_m^\pm$ are introduced in the preceding section. Hence if we know the complex amplitudes, i.e.\ the vectors $\ket{\beta_1}$ and $\ket{\beta_2}$, we can from the expression \eref{upsilon} calculate the coefficients $\upsilon_k(x)$. The result of our numerical calculations is displayed in the \Fref{fig1}.

\begin{figure}
	\centering
		\includegraphics[width=1.00\textwidth]{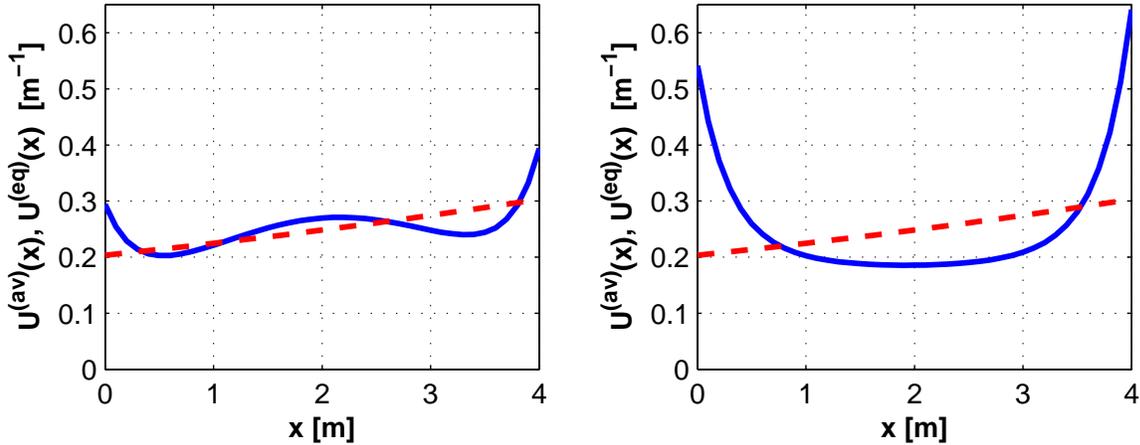}
	\caption{The comparison of time-asymptotic probability density averaged over the period of the external field $U^{(av)}(x)$ (full line) and the time-asymptotic probability density in the static case $U^{(eq)}(x)$ (dashed line). Parameters used for the calculation of the curves are in the left panel the same as used in figure~\ref{fig1}, $v_0=0.1\,\mathrm{m\,s^{-1}}$, $v_1=4.0\,\mathrm{m\,s^{-1}}$, $l=4.0\,\mathrm{m}$, $\omega=4.0\,\mathrm{rad\,s^{-1}}$ and $D=1.0\,\mathrm{m^2s^{-1}}$. The only parameter changed on the right panel is the driving frequency $\omega=2.0\,\mathrm{rad\,s^{-1}}$.}
	\label{fig2}
\end{figure}

Knowing the time-resolved probability density one could ask, how does look the average density profile. If we perform the time-averaging directly in \eref{urozklad}, we readily recognize, that the average probability density $U^{(av)}(x)$ corresponds to the zeroth Fourier coefficient $\upsilon_0(x)$. 
\begin{equation}
  U^{(av)}(x)=\frac{\omega}{2\pi}\lim_{t\to\infty}\int_t^{t+\frac{2\pi}{\omega}}\rmd t'\,U(x,t';x',0)\equiv\upsilon_0(x)\,\,.
\end{equation}
The \Fref{fig2} demonstrates the numerically calculated averaged probability density profile $U^{(a)}(x)$  and compares it with the time-asymptotic probability density $U^{(eq)}(x)$ in the static case as defined in \eref{Ueq}. For certain sets of model parameters different shapes of $U^{(a)}(x)$ can occur. Two of them are shown at \Fref{fig2}. We discuss them in the next section. 

\section{Concluding remarks}
Let now think again about the vessel in the external oscillating field that we have outlined in the beginning of the paper. We have presented rather complete picture of the long-time dynamics of the concentration profile. The solutions of the system of integral equations $B_1(t,x')$ and $B_1(t,x')$ (or their asymptotic counterparts $B_1^{(a)}(t)$ and $B_2^{(a)}(t)$) can be interpreted as a relative portion of the particles touching the respective wall of the vessel.

The space and time resolved concentration profile in our vessel can be possibly measured, e.g.\ by measuring the light absorption or transmission in the direction perpendicular to the $x$-axis. If the measuring device with higher time-resolution than the driving frequency $\omega$ would be used, one will obtain the results analogous to \Fref{fig1}, i.e.\ the function $U^{(a)}(x,t)$. On the other hand, if the time-resolution of the device would be several or more $\omega$, one can measure only the averaged concentration, i.e.\ the time-averaged density profile $U^{(av)}(x)$ demonstrated at \Fref{fig2}.

Our numerical calculation has revealed the interesting features in the behavior of the time-averaged asymptotic density $U^{(av)}(x)$. In the non-static mode, i.e.\ $v_1\neq0$ and $\omega\neq0$, the average probability density in the vicinity of the boundaries is always higher than in the static case. More precisely, $U^{(av)}(x_i)\geq U^{(eq)}(x_i)$, $i=1,2$. Differently speaking, the oscillating external field always rises the concentration at walls of the vessel. The equal sign holds in the limit $l\to\infty$. This case is equivalent to the model discussed in \cite{chvosta05b}.

The numerical results show two distinguishable shapes of the function $U^{(av)}(x)$. The realization of the particular shape is a consequence of the fine interplay of the model parameters. Both possibilities are demonstrated at \Fref{fig2}. The left panel shows the surprising situation when two local minims surround a local maximum and $U^{(av)}(x)$ crosses $U^{(eq)}$ four times. This case can only realize when the oscillating component of the field reverses the global slope of the potential during some part of its period, i.e. $v_0<v_1$ and simultaneously frequency $\omega$ is sufficiently high and the spatial interval between the boundaries is sufficiently long. In comparison with the static case, the additional oscillating field rises the average concentration in the middle-region and near the walls of the vessel and creates two regions with the lowered concentration.

The right panel of \Fref{fig2} demonstrates the situation when the driving field acts as a centrifuge. The function $U^{(av)}(x)$ crosses $U^{(eq)}(x)$ twice. The $U^{(av)}(x)$ can be both monotonically decreasing or forming the global minimum in the middle-region of the interval. The average effect of the oscillating field is the increased concentration near the walls of the vessel and diluted region in nearby its middle. Again, the realization of a particular case depends on a tuning of the model parameters.

The more detailed discussion of the asymptotic behavior of the system and of its time-averaged characteristics is obstructed by the rather complicated inversion of the matrices $\mathbb{R}_i$, $i=1,2$. It is essential during the calculation of the \emph{complex amplitudes}.

The presented mathematical procedure is not suitable only for the calculating the time resolved or time averaged probability density. Calculating the complex amplitudes, i.e. the asymptotic solution of the original system of integral equations, and going through analogous presented in Sections 4 and 5, it is also possible to check the proper normalization of $U^{(a)}(x,t)$ or to calculate the mean coordinate. Our present paper deals with the kinetic characteristic of the model. Another possibility would be to study its thermodynamics of this isothermal non-equilibrium process.

\ack{The support for this work by Ministry of Education of the Czech Republic (project no.\ MSM 0021620835) is gratefully acknowledged.}

~
%

\end{document}